\documentclass[runningheads]{llncs}
\usepackage{amsmath}
\usepackage{amsfonts}
\usepackage{amssymb}
\usepackage{graphicx}
\usepackage[utf8]{inputenc}
\usepackage[ruled]{algorithm2e}

\title{Bayesian Optimization of ESG Financial Investments}
\author{eduardogarrido90 }
\date{June 2022}

\begin{document}

\title{Bayesian Optimization of ESG Financial Investments}
\titlerunning{Bayesian Optimization of ESG Financial Investments}
\author{Eduardo C. Garrido-Merchán, Gabriel González Piris, Maria Coronado Vaca}
\date{June 2022}

\institute{Universidad Pontificia de Comillas, Madrid, Spain \\
\email{ecgarrido@icade.comillas.edu, gabrielgonzalezpiris@alu.icai.comillas.edu, mcoronado@icade.comillas.edu}}

\maketitle

\begin{abstract}
Financial experts and analysts seek to predict the variability of financial markets. In particular, the correct prediction of this variability ensures investors successful investments. However, there has been a big trend in finance in the last years, which are the ESG criteria. Concretely, ESG (Economic, Social and Governance) criteria have become more significant in finance due to the growing importance of investments being socially responsible, and because of the financial impact companies suffer when not complying with them. Consequently, creating a stock portfolio should not only take into account its performance but compliance with ESG criteria. Hence, this paper combines mathematical modelling, with ESG and finance. In more detail, we use Bayesian optimization (BO), a sequential state-of-the-art design strategy to optimize black-boxes with unknown analytical and costly-to compute expressions, to maximize the performance of a stock portfolio under the presence of ESG criteria soft constraints incorporated to the objective function. In an illustrative experiment, we use the Sharpe ratio, that takes into consideration the portfolio’s returns and its variance, in other words, it balances the trade-off between maximizing returns and minimizing risks. On the other hand, ESG considers a wide variety of criteria such as climate change, business ethics or corporate governance. In the present work, ESG criteria have been divided into fourteen independent categories used in a linear combination to estimate a firm’s total ESG score. Most importantly, our presented approach would scale to alternative black-box methods of estimating the performance and ESG compliance of the stock portfolio. In particular, this research has opened the door to many new research lines, as it has proved that a portfolio can be optimized using a BO that takes into consideration financial performance and the accomplishment of ESG criteria. 
\end{abstract}


\section{Introduction}
In the past 15 years, ESG criteria have increasingly become integrated into mainstream portfolio management. ESG investment -also called socially responsible investment (SRI)- has attracted much attention from both institutional and individual investors in capital markets \cite{fan2020sustainable,gibson2022responsible,chen2021social,gutsche2019private,oikonomou2018socially,melas2017factor,hitchens2015finding,bollen2007mutual}. Many long-term institutions such as pension funds, insurance companies, sovereign wealth funds, foundations and endowments have signed up to the UN Principles for Responsible Investment (PRI). The six PRI offer a menu of possible actions for incorporating ESG issues into investment practice. The PRI were developed in 2006 by an international group of institutional investors reflecting the increasing relevance of environmental, social and corporate governance (ESG) issues to investment practices. The process was convened by the United Nations Secretary-General. As of December 2022, the PRI has 5,179 signatories, representing US\$121 trillion of assets under management (AUM) -a huge increase from US\$6.5 trillion in 2006- (UNPRI, 2022). In addition, individual investors show increasingly strong activism claiming for their money being invested in ESG assets and PRI signatories rely on shareholder activism to pursue responsible investing \cite{anderson2012will}.

In addition, the latest United Nations Global Compact-Accenture CEO Study on Sustainability (UN, 2023) provides insights and a resolute call to action from a record from over 2,600 CEOs across 128 countries and 18 industries. Sustainability has emerged as the core of resilience. Compared to the 2013 CEO Study, CEOs now unequivocally feel it is their role to make their business more sustainable (98\% agree versus 83\% in the 2013 edition). Sustainability is the only path to build a truly resilient company. While CEOs remain committed to the UN Sustainable Development Goals (SDGs), they’re sounding the alarm that more action is needed to rescue the goals. To remove barriers on taking action, CEOs are calling for a new roadmap to achieve the SDGs and asking government to accelerate the green transition. It also comes at a critical time. As we approach the halfway mark to achieve the UN Sustainable Development Goals (SDGs) by 2030, the message from CEOs is painfully clear: we must accelerate our work in sustainability to build a more resilient future if we are to rescue the SDGs by 2030 (UN, 2023) The 2021 CEO Study found that business leaders were severely off track to deliver on their sustainability and climate goals. Today, the situation is even more tenuous. Facing continued fallout from the pandemic, coupled with the effects of Russia’s war in Ukraine, broader geopolitical uncertainty, inequality, and climate change, CEOs report heightened frustration and uncertainty in preparing for what will happen next. As a result, CEOs are now making sustainability a top priority in their agendas, re-evaluating their investment criteria and developing innovative business models enabled by technology to drive change (UN, 2023). 

Thus, ESG criteria have become more significant due to the growing importance of investments being socially responsible, and because of the financial impact companies suffer when not complying with ESG criteria. The E of Environmental takes into account the direct and indirect environmental impact of the company’s activities, the S of Social considers the impact of the company in society and its community, and the G of Governance takes into account the corporate governance of the company.  

The interest of investors towards ESG funds and assets has grown lately. Bank of America Global Research states in its report “EDFR Global” that the flow of money into ESG equity funds has doubled from 2020 to 2021. Additionally, it is strongly believed by many investors that integrating ESG criteria when investing increases returns and profitability. The global asset management firm Macquarie AM states in its report “ESG survey” that three out of four institutional investors consider that taking into account ESG criteria in its investments increases financial returns.  Empirical evidence in this respect is mixed as several studies show the existence of a negative relationship between the environmental and financial performance of portfolios while others argue in favor of a positive effect. For this, we refer to the most recent studies and meta-analysis conducted on this area \cite{atz2022does,whelan2021esg}. 

As a result of the situation described above, the relevance of ESG investing involves the need of more research on the alternative portfolio optimization techniques within the ESG framework. Despite the voluminous literature that analyzes the financial performance of ESG portfolios compared to "conventional” -or non-ESG- ones, we believe that the portfolio optimization frameworks used to determine the asset weights in ESG portfolios is underdeveloped. 

Moreover, Oikonomou et al. (2018) proved that different optimization techniques lead to different ESG portfolio performance \cite{oikonomou2018socially}. Hence, apart from the ESG screening criteria, investors and portfolio managers also need to carefully consider the choice of asset allocation method. So, it is claimed to consider and investigate alternative optimization techniques within the ESG framework.  Different models for portfolio optimization have been applied to ESG portfolios but no previous studies have applied BO to ESG portfolios. This being the research gap to which the present paper responds. The present study is the first report (to the best of our knowledge) that leverages Bayesian optimization and ESG scores into a portfolio optimization model. This paper aims to analyze whether BO can be applied within the ESG portfolio optimization framework as an alternative to models such as the Markowitz one when its assumptions fail. To show the empirical performance of BO, we carry out two illustrative experiments. Our main contribution is to show that BO can be applied within the ESG portfolio optimization framework. In particular, this paper seeks to contribute to the literature through investigating whether BO can be used in the world of ESG investing, opening a further research line in portfolio allocation.  

This paper is organized as follows. First, we begin with a state-of-the-art description about ESG portfolio optimization and Bayesian optimization methods. Then, we describe the components of the objective function that we will use in the illustrative experiment. In a following section, we describe Bayesian optimization in detail in order to understand the illustrative experiment section, that comes afterwards. Finally, we include in the manuscript a conclusions and further work section that closes our work.  

\section{State of the art in portfolio optimization with ESG and Bayesian optimization}

Different traditional and widely known models for portfolio optimization have been applied to ESG portfolios \cite{oikonomou2018socially} such as the Markowitz mean-variance portfolio optimization approach \cite{markowitz1952portfolio}, or the Black and Litterman asset allocation model \cite{black1992global}. Likewise, more recent portfolio optimization methods or with a less solid mathematical basis have been used for the construction of ESG portfolios (the naïve diversification approach - or 1/N portfolio strategy- \cite{oikonomou2018socially}, the risk-parity portfolio framework (\cite{anderson2012will,baker2011benchmarks}) and the reward-to-risk timing portfolio strategy \cite{kirby2012s}.

Markowitz mean-variance portfolio optimization model only considers risk and return and does not allow for additional criteria. The need for portfolio selection to be able to include criteria beyond mean and variance is solved with multi-criteria portfolio selection. Many multiple criteria methods have already been applied in the field of portfolio selection since Lee \& Lerro \cite{lee1973optimizing}; we refer to Aouni et al. (2018) for a review \cite{aouni2018increasing}. Concerning studies that apply multi-criteria methods to optimize ESG portfolios we can cite, among others, the following: managing ESG portfolios from a linear multicriteria approach \cite{hallerbach2004framework}. A multicriteria approach but in a classical utility theory under uncertainty framework, instead of a linear one \cite{ballestero2012socially}. A two-stage multi-objective framework for the selection of ESG portfolios by applying a “Hedonic Price Method”, selecting ESG portfolios using goal programming models and fuzzy technology \cite{bilbao2012selection}. A model that combines goal programming with “goal games” against nature \cite{trenado2014corporate}. A tri-criterion framework for inverse optimization of ESG portfolios \cite{utz2014tri}. An integration of the ESG portfolio selection problem into a Decision Support System \cite{calvo2015finding}. A multicriteria portfolio selection model for mutual funds based on the Reference Point Method \cite{mendez2015mutual}.  Similarly, a Markowitz’ model modification through a new tri-criterion model enabling investors to custom–tailor their asset allocations and incorporate all personal preferences regarding return, risk and social responsibility \cite{gasser2017markowitz}. Additionally, a formulation of the portfolio optimization problem as a multiple-objective problem, where the third objective corresponds to corporate social responsibility \cite{qi2018outperforming}. Analogously, three adaptations of multiobjective evolutionary algorithms to include a social screening preceding the optimization process \cite{liagkouras2020incorporating}. Finally, a hybrid ESG portfolio selection model with multi-criteria decision making (MCDM) and multi-objective optimization problem (MOOP) techniques \cite{wu2022integrated}. 

Other authors propose ESG-adjusted capital asset pricing models: the Sustainable CAPM model (S-CAPM) \cite{pedersen2021responsible,zerbib2022sustainable}. This leads us to ESG factor models or the ESG factor investing strand of literature which considers ESG criteria as a traditional systematic risk factor, either as a standalone factor or as a subcomponent of factor strategies \cite{kaiser2020esg,halbritter2015wages,fan2020sustainable,derwall2005eco,melas2017factor,gorgen2020carima,roncalli2020measuring,alessi2021greenium,gimeno2022role}. Additionally, a data envelopment analysis (DEA) model with quadratic and cubic terms to enhance the evidence of two or more aspects, as well as the interaction between the environmental, social, and governance attributes has been proposed \cite{chen2021social}. They then combined the ESG scores with financial indicators to select assets based on a cross-efficiency analysis. 

It has been defended that traditional portfolio optimization methods are inadequate for this kind of investments. Consequentially, a fusion approach of machine learning and portfolio selection for ESG investments has been proposed \cite{vo2019deep}. In particular, they proposed a deep reinforcement learning model – that contains a Multivariate Bidirectional Long Short-Term Memory (LSTM) neural network – to predict stock returns for the construction of a ESG portfolio. They called their new model Deep Responsible Investment Portfolio (DRIP). 

Thus, different models for portfolio optimization have been applied to ESG portfolios by few studies but no previous research has applied BO to ESG portfolios. If the assumptions done by other models such as the Markowitz model are not true about a particular market, their results will not be reliable. Hence, the portfolio return will be a black-box. Consequently, this is the research gap to which the present paper responds. The present study is the first report (to the best of our knowledge) that leverages Bayesian optimization and ESG scores into a portfolio optimization model. This paper seeks to contribute to this underdeveloped strand of literature through investigating whether BO can be used in the world of ESG investing.  

\section{The objective function: Sharpe Ratio subject to ESG Criteria}

In this section we will describe and motivate our objective function. In particular, we will minimize the Sharpe ratio of a portfolio under the presence of soft constraints consisting of ESG criteria. As we will further see, we include the ESG soft constraints in the objective function, acting as penalization criteria for the objective function. Interestingly, it is useful to model the optimization problem like this as these constraints are not hard, in the sense that they penalize the possible solutions but do not make them unfeasible. The following subsections describe the two components of the objective function: the ESG criteria and the Sharpe ratio. After describing them, we explain how do we integrate them in the objective function.  

\subsection{Fundamentals of ESG Criteria}
ESG criteria is a framework for analysing and assessing an organization’s performance in environmental, social and governance matters in comparison with its competitors \cite{hvidkjaer2017esg}. 

\begin{figure}[h]
    \centering
    \includegraphics[width = 0.8\textwidth]{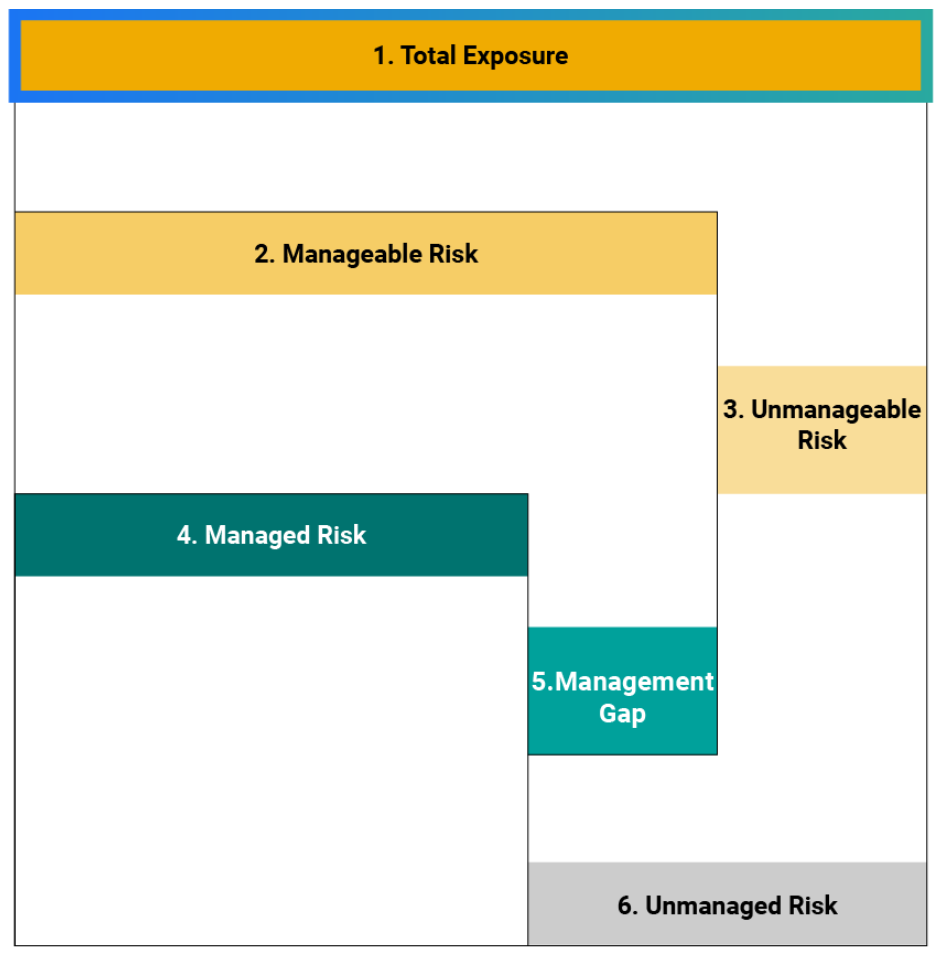}
\caption{ESG Criteria Risks.}
    \label{fig:im1}
\end{figure}

The six most popular/prominent ESG rating agencies are Sustainalytics ESG Risk Rating, MSCI ESG Ratings, Moody´s ESG (formerly Vigeo-Eiris), Refinitiv (formerly Asset4), Bloomberg ESG Disclosures Scores, and S\&P Global ESG Scores (formerly RobecoSAM). These agencies belong to some of the largest financial groups, such as Morningstar, MSCI or Bloomberg. The scores given to the different companies are industry oriented, which means they must be compared with their industry competitors in order to evaluate their compliance with ESG criteria. Some rating agencies, such as Sustainalytics ESG Risk Rating, divide the risk of not meeting with ESG criteria into manageable and unmanageable risks, as it can be seen in Figure \ref{fig:im1}. Manageable risks are then divided into managed risks and management gap. Management gap represents the potential improvement in compliance with ESG criteria companies have.  

Rating agencies look mainly at different aspects or categories in each of the three key ESG areas when evaluating a company's ESG performance, such as the following ones: 

Regarding the environmental category, we find climate change: the impact of climate change in a company is assessed by looking at different criteria such as the Carbon Footprint of their products and organisation, carbon and greenhouse emissions and their vulnerability to climate change. Natural capital: the use and dependency of a company to natural capital depends on different factors such as water stress, its use of land or its use of natural resources. Pollution and waste: In this category aspects such as toxic emissions, the use of packaging material and electronic waste are considered. Environmental opportunities: includes issues such as using renewable energy or green building.

Considering the social category, we have the human capital: this category includes several aspects such as health and safety of the workforce or labour management. Product liability: chemical safety, product safety and quality, or privacy and consumer protection belong to this category. Stakeholder opposition: This section includes opposition of stakeholders, such as local communities or suppliers, to a company. Social opportunities: this group includes several aspects such as access to healthcare or communication.  

Finally, regarding governance, we include corporate governance: This category includes a variety of aspects such as management structure or executive compensation. Corporate behaviour: topics such as business ethics or transparency are included in this section. Business transparency refers to the access to quality corporate information from all stakeholders and the extent to which its actions are observable by stakeholders. In the European Union, the Non-Financial Reporting Directive (NFRD) (Directive 2014/95/EU) regulates and establishes the extent to which companies in the EU must comply with corporate transparency. It also establishes that all EU corporations must include non-financial statements in their annual reports.  

\begin{figure}[h]
    \centering
    \includegraphics[width = 0.99\textwidth]{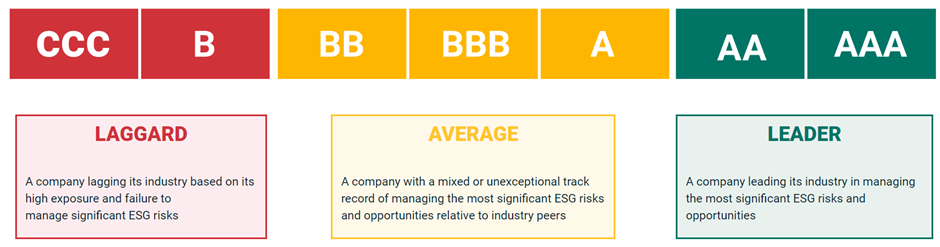}
\caption{MSCI ESG Score.}
    \label{fig:im2}
\end{figure}

Rating agencies give scores in different metrics to represent a company’s overall ESG risk. Just as an illustrative example, MSCI gives a score between CCC and AAA, as it can be seen in Figure \ref{fig:im2}.

However, other agencies as Bloomberg give a [1-100] scale or Sustainalytics uses a 5 risk level range according to a [0-40] score from 10 to 10. This is because there is no benchmark to which comparing the accomplishment of the different criteria. Therefore, a company’s ESG score doesn’t provide much information about a firm’s environmental, social and governance performance unless it is compared with the scores of other companies in the same sector or industry. Additionally, it is important to compare ESG scores of firms in the same sector, not across sectors, as there are industry related factors that affect a firms ESG performance. For example, a company in an industry which requires high amounts of energy resources, such as a chemical company, is not comparable with a service company, such as a consultancy firm.  

One of the criticisms made to ESG criteria derives from the fact that they fall short in taking into consideration the overall mission of corporations. Additionally, some of the criteria may be considered to be subjective, as analysts will have to score the companies according to their views and opinions on the performance on those particular areas. Therefore, in order to provide an ESG metric that encodes all, or the majority, of the mentioned criteria, it is important to consider different ESG scores and approaches when assessing a firm’s performance in the three areas. For our work, as an illustrative example, we used the following ESG score, but we emphasize that our methodology is compatible with any subjective or objective ESG score, as we use Bayesian optimization that is able to deal with black-box functions whose gradients are unknown and whose values can be subjective.  

\begin{figure}[h]
    \centering
    \includegraphics[width = 0.99\textwidth]{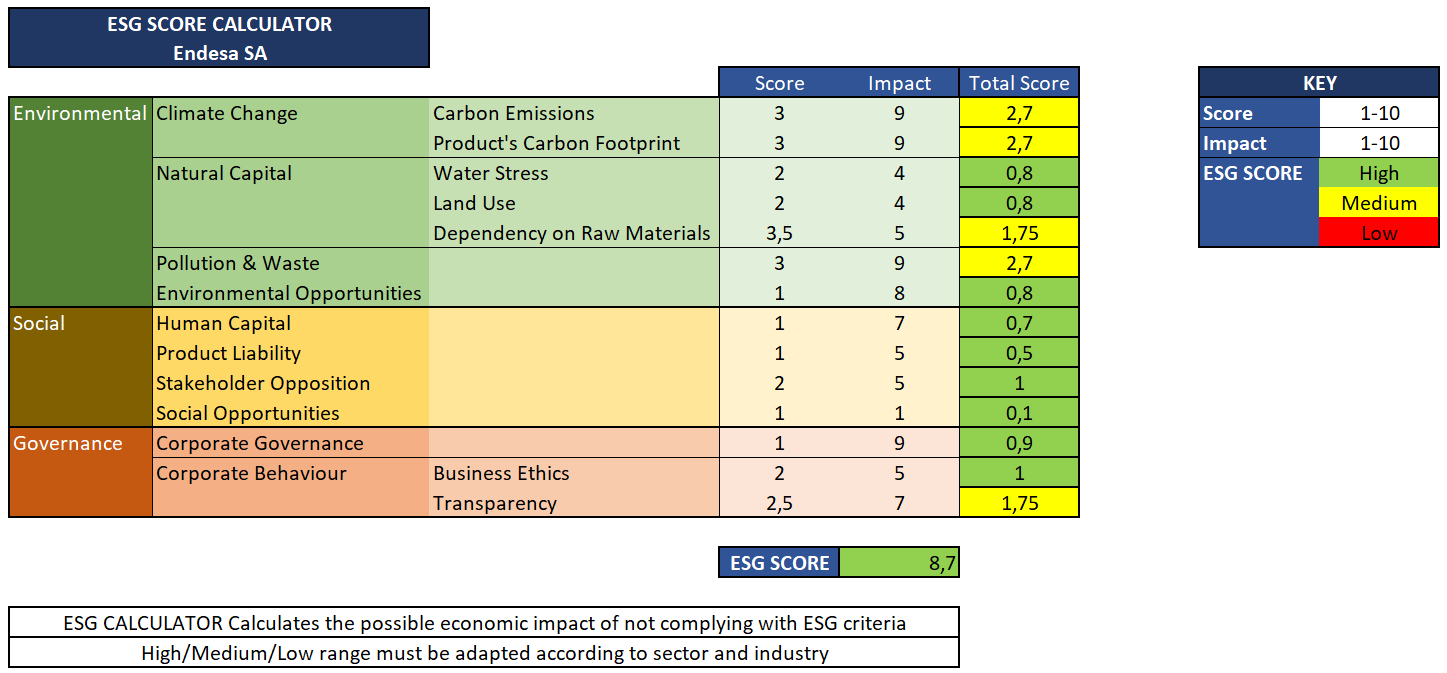}
\caption{Example of the proposed ESG score in the Endesa company.}
    \label{fig:im3}
\end{figure}

In Figure \ref{fig:im3} the ESG Score of Endesa can be seen. Endesa received a score of 8,7, and had no category classified as “low”. The category where Endesa performs best according to ESG criteria is Product Liability, followed by Human Capital. Meanwhile, the categories where it performed worst are Carbon Emissions and Product’s Carbon Footprint. This is because, Endesa still has thermal plants operating, which are very pollutant, as they consume carbon. Additionally, it has combined cycles, which also emit greenhouse gases, although less than thermal plants, as they consume natural gas. Endesa’s ESG score is expected to improve in the following years, as it plans to close all its carbon businesses by 2027 and its combined cycles in 2040. 

\subsection{Sharpe Ratio}
The Sharpe ratio takes into consideration an asset’s return and its variance \cite{sharpe1994sharpe}. Therefore, it balances the trade-off between maximizing returns and minimizing the risk or volatility.  

Equation 1 shows the Sharpe Ratio used in the present research. The Sharpe ratio takes into consideration in the numerator the return and weight of the different assets in the portfolio and the risk-free rate. In the denominator it considers the covariance matrix of the portfolio and the weight of the different assets. The diagonal of the covariance matrix is the variance of the different assets. The covariance matrix is symmetric about the diagonal.  

\begin{align}
    f_x = \frac{\sum_{i=1}^{N} w_i r_i - r_f}{\sigma_p} \quad s.t. \sum_{i=1}^{N} w_i = 1, 0\leq w_i \leq 1,
\end{align}

where $N$ is the number of different assets, $w_i$ is the weight of each asset $i$ in the portfolio, $r_i$ is the return of asset i, $r_f$ is the risk-free rate and $\sigma_p$ is the standard deviation of the excess return of the portfolio. 

\subsection{ESG-constrained Sharpe Ratio}
\begin{figure}[h]
    \centering
    \includegraphics[width = 0.8\textwidth]{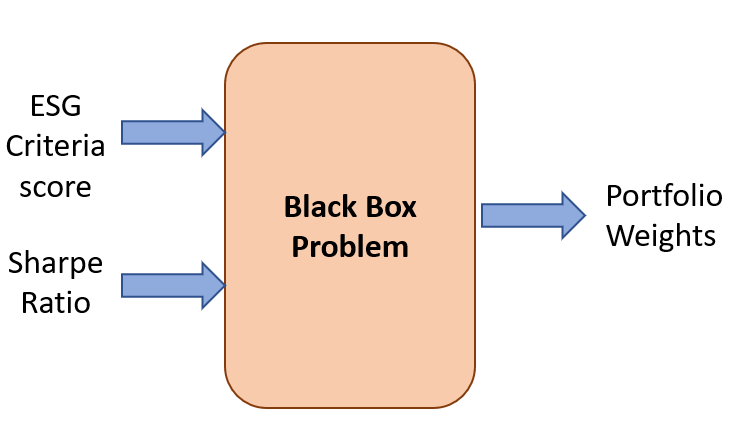}
\caption{ESG objective function of the ESG Portfolio optimization.}
    \label{fig:im4}
\end{figure}

We combine the previous two concepts –ESG criteria and Sharpe ratio- in a single objective function that we will optimize with the method that will be illustrated in the following section, the bayesian approach. In particular, the objective function tries to optimize the Sharpe ratio being penalized by the ESG criteria. The optimization of the objective function, being analytical or a potential black-box, will return the weights of the optimal portfolio in terms of performance and ESG compliance as Figure \ref{fig:im4} illustrates. 

Combining both factors requires them to have the same magnitude in the objective function. For this purpose, they are normalized. The ESG score is normalized taking into consideration its maximum and minimum values which are 0 and 10. Analogously, the Sharpe ratio is also normalized using its maximum and minimum values. In particular, the maximum and minimum values are estimated by their sample values on the dataset commented in the illustrative experiment. Then, we simply add both factors. However, the ESG factor can be added a logarithm factor and normalized again if we have low ESG scores to highly penalize the objective function and higher ESG scores to make the objective only slightly worse. As the ESG score normalized and the Sharpe ratio normalized take values in the interval between 0 and 1, the fitness function takes values in the interval between 0 and 2. 

\section{Fundamentals of Bayesian optimization}
Bayesian optimization is a state-of-the-art class of methods that optimize black-boxes, that is, unknown noisy analytical functions that are very expensive to evaluate whether in time or computational resources \cite{garrido2021advanced}. For example, the estimation of the generalization error of machine learning algorithms with respect to their hyper-parameters is considered to be a black-box function and the first successful application of Bayesian optimization (BO) \cite{snoek2012practical}. In order to solve such a scenario, we need a method coping with the optimization of a black-box without using gradients, in a small number of steps, and considering noise in the evaluations. More formally, the purpose of BO is to retrieve the optimum $\mathbf{x}^\star$ of a black-box function $f(\mathbf{x})$ where $\mathbf{x} \in \mathcal{X}$ and $\mathcal{X}$ is the input space where $f(\mathbf{x})$ can be observed. In other words, we want to retrieve $\mathbf{x}^\star$ such that,
\begin{equation}
\mathbf{x}^\star = \arg\min_{\mathbf{x} \in \mathcal{X}}f(\mathbf{x})\,,
\end{equation}
assuming minimization. We can define a BO method by the following tuple
\begin{equation}
\mathcal{A} = (\mathcal{M}, \alpha(\cdot), p(f(\mathbf{x})|\; \mathcal{D}))\,,
\end{equation}
where $f(\mathbf{x})$ is the black-box that we want to optimize, $\mathcal{M}$ is a probabilistic surrogate model, $\alpha(\cdot)$ is an acquisition, or decision, function, $p(f(\mathbf{x})|\mathcal{D})$ is a predictive distribution of the observation of $\mathbf{x}$ and $\mathcal{D} = \{(\mathbf{x}_i, y_i)| i = 1,...,t\}$ is the dataset of previous observations at iteration $t$.

To successfully solve this task, BO uses a probabilistic surrogate model, being a common option the Gaussian process (GP), of the target function. Concretely, a GP is a set of random variables (of potentially infinite size), any finite number of which have (consistent) joint Gaussian distributions \cite{williams2006gaussian}. More formally, a GP is fully defined by a zero mean and a covariance function or kernel $k(\mathbf{x},\mathbf{x}')$, that is, $f(\mathbf{x}) \sim \mathcal{G}\mathcal{P}(\mathbf{0},k(\mathbf{x},\mathbf{x}'))$. More concretely, the covariance function of the GP receives two points as an input, $\mathbf{x}$ and $\mathbf{x}'$. Given a set of observed data $\mathcal{D} = \{(\mathbf{x}_i, y_i)| i = 1,...,N\}$, where $y_i=f(\mathbf{x}_i) + \epsilon_i$ with $\epsilon_i$ some
additive Gaussian noise, a GP builds a Gaussian predictive distribution $p(f(\mathbf{x}^\star)|\mathcal{D})$ for the potential values of $f(\mathbf{x}^\star)$ at a new input point $\mathbf{x}^\star$. Concretely,
$p(f(\mathbf{x}^\star)|\mathcal{D}) =\mathcal{N}(f(\mathbf{x}^\star)|
\mu(\mathbf{x}^\star),  v(\mathbf{x}^\star))$. Lastly, the mean $\mu(\mathbf{x}^\star)$ and variance $v(\mathbf{x}^\star)$ of the predictive distribution $p(f(\mathbf{x}^\star)|\mathcal{D})$ are respectively given by:
\begin{align}
\mu(\mathbf{x}^\star) & = \mathbf{k}_{\star}^{T} (\mathbf{K}+\sigma^{2}\mathbf{I})^{-1}\mathbf{y}\,, \\
v(\mathbf{x}^\star) & = k(\mathbf{x}_{\star},\mathbf{x}_{\star}) - \mathbf{k}_{\star}^T(\mathbf{K}+\sigma^{2} \mathbf{I})^{-1}\mathbf{k}_\star\,,
\label{eq:pred_dist}
\end{align}
where $\mathbf{y}=(y_1,\ldots,y_N)^\text{T}$ are the observations collected so far;
$\sigma^2$ is the variance of the additive Gaussian noise $\epsilon_i$;
$\mathbf{k}_\star = \mathbf{k}(\mathbf{x}_*)$ is a $N$-dimensional vector with the prior covariances between the test point $f(\mathbf{x}^\star)$ and
each of the training points $f(\mathbf{x}_i)$; and $\mathbf{K}$ is a $N\times N$ matrix with the prior covariances among each $f(\mathbf{x}_i)$, for $i=1,\ldots,N$. Each element $\mathbf{K}_{ij} = k(\mathbf{x}_i, \mathbf{x}_j)$ of the matrix $\mathbf{K}$ is given by the covariance function between each of the training points $\mathbf{x}_i$ and $\mathbf{x}_j$ where $i,j = 1,...,N$ and $N$ is the total number of training points.

Using this model, we can estimate a predictive distribution of the unknown function in areas of the space where it has not been evaluated yet. Using this distribution, BO computes, iteratively, an acquisition function $\alpha(\mathbf{x})$ that estimates, for every input space point $\mathbf{x}$, the expected utility of evaluating the objective. In particular, the point whose value maximizes the acquisition function is suggested for evaluation in an iterative fashion. Most critically, that point maximizes the compromise between exploration of unknown areas and exploitation of promising solutions evaluated before. The acquisition function $\alpha(\mathbf{x})$ is generally not difficult to maximize. In particular, we can compute the gradient $\nabla_\mathbf{x} \alpha(\mathbf{x})$ of the acquisition function and use it for its optimization. We can compute the gradient $\nabla_\mathbf{x} \alpha(\mathbf{x})$ because the acquisition function $\alpha(\mathbf{x})$ is cheap to evaluate, as it is only based on the GP predictive distribution $p(f(\mathbf{x})|\mathcal{D}))$. An example of acquisition function is the expected improvement (EI). Let $\chi(\mathbf{x}) = \frac{\mu(\mathbf{x}) - \kappa - \epsilon}{\sigma(\mathbf{x})}$, EI is given by:
\begin{equation}
\text{EI}(\mathbf{x}) = (\mu(\mathbf{x}) - \kappa - \epsilon) \Phi(\chi(\mathbf{x})) + \sigma(\chi(\mathbf{x})) \phi(\chi(\mathbf{x}))\,,
\end{equation}
where $\Phi(\cdot)$ is the Gaussian CDF and $\phi(\cdot)$ is the Gaussian PDF. Concretely, it is a heavily based exploitative criterion.

Afterwards, the GP is updated with the suggestion and conditioned there, obtaining a new predictive distribution and making the BO method repeat the described instructions iteratively, until a number of evaluations is consumed, where BO gives the final recommendation. In particular, this point can be the one whose evaluation has the best observed value or the point that optimizes the GP predictive mean. We summarize the steps of the basic Bayesian optimization method in the following algorithm.

\begin{figure}[tb]
\begin{algorithm}[H]
\label{alg:bo}
\textbf{Input:} Maximum number of evaluations (budget) $T$.
\caption{BO of a black-box function $f(\mathbf{x})$.}
\For{$\text{t}=1,2,3,\ldots,T$}{
        {\bf 1:} {\bf if $N=1$:}\\
        \hspace{.7cm}Choose $\mathbf{x}_t$ randomly from $\mathcal{X}$. \\
        \hspace{.5cm}{\bf else:} \\
        \hspace{.7cm}Find $\mathbf{x}_t$ by optimizing the acquisition function:
        $\mathbf{x}_t = \underset{\mathbf{x} \in \mathcal{X}}{\text{arg max}} \quad \alpha_t(\mathbf{x})$.

        {\bf 2:} Evaluate the black-box function $f(\cdot)$ at $\mathbf{x}_t$: $y_t=f(\mathbf{x}_\text{t}) + \epsilon_t$.

        {\bf 3:} Augment the dataset with the observed value: $\mathcal{D}_{1:t}=\mathcal{D}_{1:t-1} \bigcup \{\mathbf{x}_t, y_t\}$.

        {\bf 4:} Fit again the GP model using the augmented dataset $\mathcal{D}_{1:t}$.
 }
{\bf 5:} Obtain the recommendation $\mathbf{x}^\star$ (best observed value). \\
\KwResult{Recommended point $\mathbf{x}^\star$}
\vspace{.5cm}
\end{algorithm}
\end{figure}

\begin{figure}[h]
    \centering
    \includegraphics[width = 0.8\textwidth]{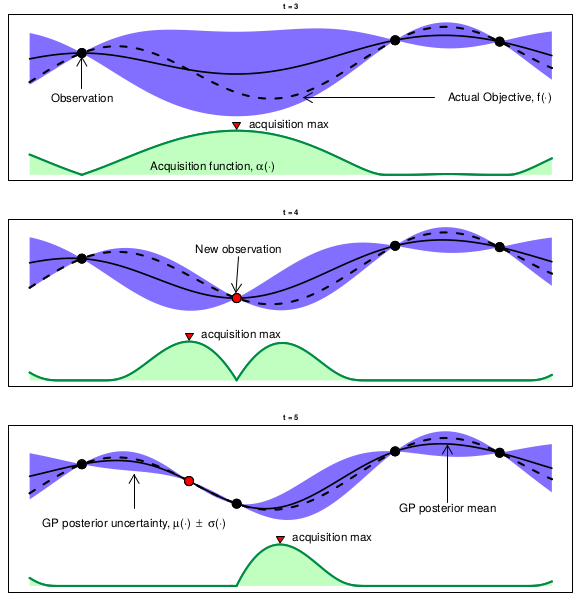}
\caption{Bayesian optimization algorithm used to obtain an optimal solution of an unknown target function \cite{garrido2021advanced}. Best seen in color.}
    \label{fig:im5}
\end{figure}

We illustrate the described process on Figure \ref{fig:im5}. In particular, this figure is divided into 3 plots, that are 3 iterations of the BO algorithm. Plotted on dotted black is the unknown objective function that we want to optimize. We also plot in black the predictive mean of the conditioned Gaussian process on the previous observations of the objective (black dots). The blue clouds are the uncertainty coming from the GP predictive distribution. Please observe how the uncertainty grows in unexplored areas of the space. Additionally, we plot as a red dot the suggestion of the Bayesian optimization algorithm. This observation comes at the exact point of the previous maximization of the acquisition function, that is represented as a green area. Recall that the acquisition function is easy to maximize, for example by a quasi-Newton second order method as the L-BFGS algorithm, as we have gradients. The represented process continues in an iterative fashion until a budget of operations is consumed, where the recommended solution comes from the minimization of the predictive mean of the Gaussian process. 

Concretely, different acquisition functions, like the expected improvement or the upper confidence bound can be used. However, they are all trade-offs between exploration and exploitation. Analogously, different probabilistic surrogate models, like random forests or Bayesian neural networks can also be used. For more information about Bayesian optimization and its applications we conclude giving a reference to study this class of methods in more detail \cite{shahriari2015taking}. 

\section{Illustrative experiments}
We use the the Bayesian optimization method described in section 4 to optimize the ESG constrained Sharpe ratio objective function presented in section 3.3. However, for illustrative purposes, we use a function whose expression is known in this illustrative experiment. In a real-world scenario the expression for the performance of the stock portfolio may be only obtained by a Monte Carlo simulation of some events whose expression is unknown and the ESG score may be obtained by a webcrawler of the different companies at the current day of the simulation, which is also an unknown expression. Hence, only Bayesian optimization could be used in the described setting as a solution, as the analytical expression is unknown and costly.  
The illustrative experiment performed in this research consists of the optimization of a portfolio containing three companies from the Spanish electric utility and energy sector. These companies are: Endesa, Repsol and Iberdrola. The three companies compete on the electric utility sector, although there are some key differences between them, which will be analysed. Repsol has a strong presence in the oil and gas sector, therefore it should have a lower ESG score, and the optimization should penalize Repsol with respect to the other two companies. The optimization algorithm considers two factors: the current ESG score of the three companies and their return in the last 12 months, from March 15th, 2021 to March 14th, 2022. We begin the section by analyzing its ESG score. We emphasize the notice already pointed out above: that the ESG scores cannot be computed using an analytical-form expression in other scenarios.  

\subsection{ESG analysis of the companies}
\begin{figure}[h]
    \centering
    \includegraphics[width = 0.99\textwidth]{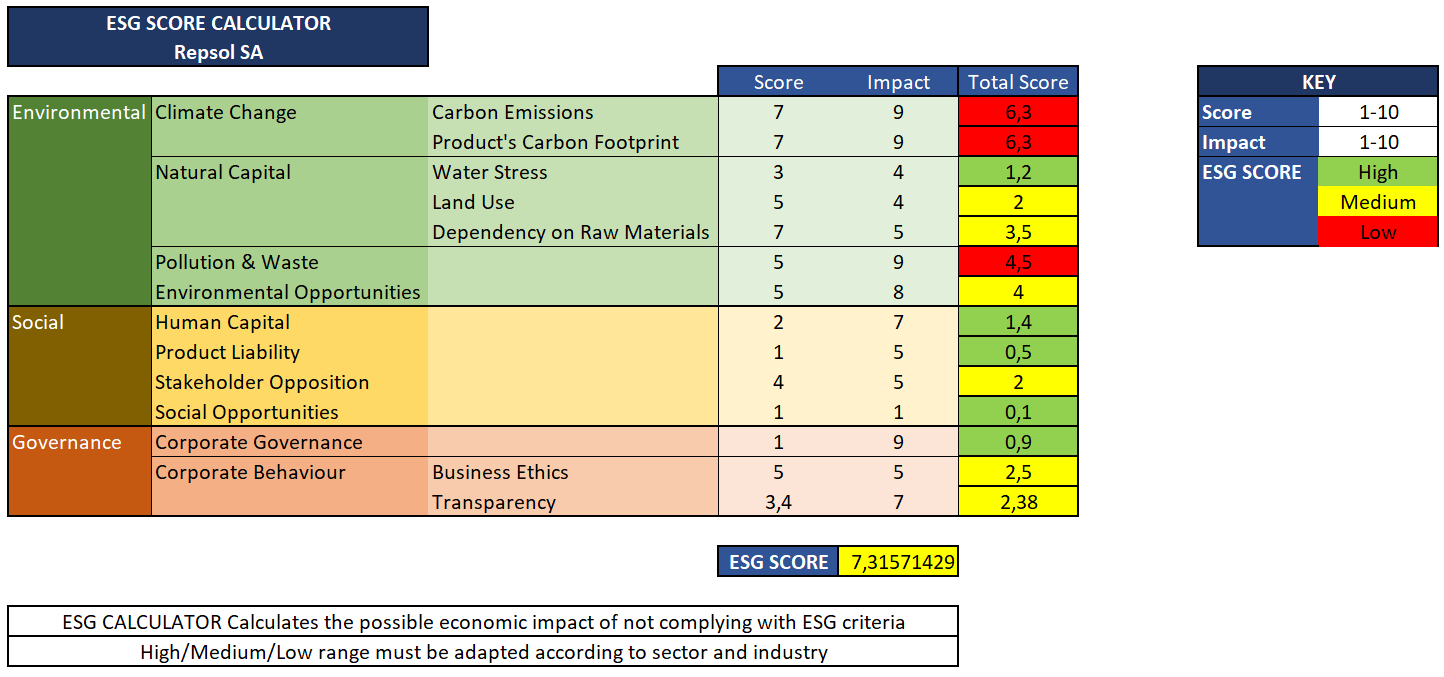}
\caption{Example of the chosen criteria of the proposed ESG score in the Repsol company.}
    \label{fig:im6}
\end{figure}

Endesa, as seen in Figure \ref{fig:im6} and described in section 3.1, had an ESG score of 8,7. As it can be seen in Figure \ref{fig:im6}, Repsol received an ESG score of 7,3, having the worst score out of the three companies. Repsol has three categories that were classified as having a “low” ESG score, which were Carbon Emissions, Product’s Carbon Footprint and Pollution \& Waste. This makes sense, as Repsol’s main business is the extraction, refinement and commercialization of hydrocarbons. The category in which Repsol performed best was Product Liability. Nevertheless, Repsol had some categories classified as “Medium” in Social and in Governance. These classifications have nothing to do with operating in the oil and gas sector, therefore it has a considerable margin of improvement in these categories. Repsol’s ESG score is also expected to improve in the coming years, as it plans to become carbon neutral in 2050. 

\begin{figure}[h]
    \centering
    \includegraphics[width = 0.99\textwidth]{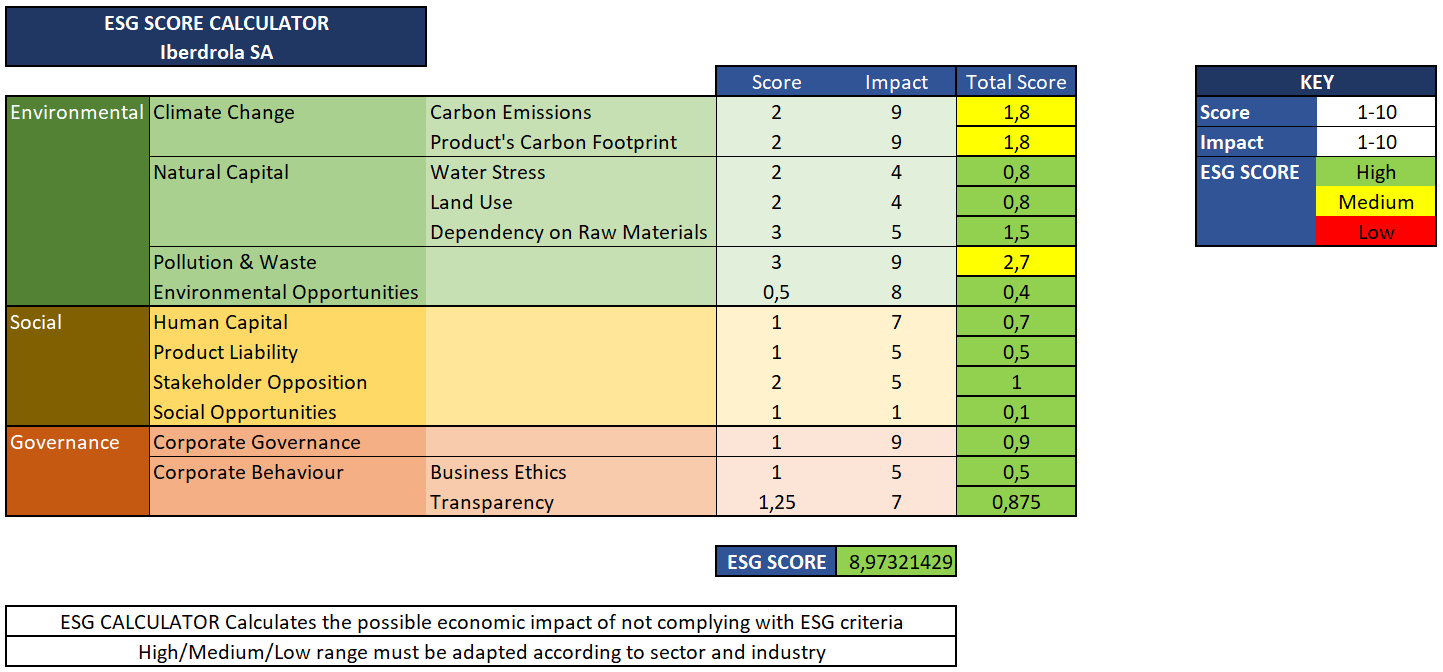}
\caption{Example of the chosen criteria of the proposed ESG score in the Iberdrola company.}
    \label{fig:im7}
\end{figure}

In Figure \ref{fig:im7} the ESG score of Iberdrola can be seen. Iberdrola received a score of 8,97, being the company with the highest score. It had no category classified as “low” and only three categories classified as “medium”. The categories in which Iberdrola performed best are Environmental Opportunities, Product Liability and Business Ethics. Iberdrola received an excellent score in Environmental Opportunities as it is the company which is planning to become carbon neutral earlier, in 2030, and is the one investing heavier in renewable energy. The categories in which Iberdrola performed worst are Carbon Emissions and Product’s Carbon Footprint, as Iberdrola still has combined cycles operating and commercializes electricity produced from all types of technologies, including thermal plants. Iberdrola performed better than Endesa in these categories, as it has a larger proportion of renewable energy resources and because it has no thermal plant operating. Iberdrola’s ESG score is expected to improve in the following years as it becomes carbon neutral. 

\subsection{ESG constrained optimization of the Sharpe Ratio}
Having computed the ESG scores, we now combine the companies and optimize the Sharpe ratio of the stock portfolio with respect to the participation of every company in the portfolio but constrained to the ESG scores. Concretely, the risk-free rate chosen for this experiment is 1.2\% which was the average yield of the Spanish 10-year treasury bond during March 2022.  

To normalize the Sharpe ratio, the maximum and minimum values need to be calculated. The formula used for the Sharpe ratio is represented and explained in section 3.3. To find the maximum and minimum value for the Sharpe ratio, the average returns and covariances for the three companies in the selected time period need to be computed. For Endesa we had an average return of -0.051\%, for Repsol we had an average return of 0.036\% and for Iberdrola we had an average return of -0.022\%.

The maximum value of the Sharpe ratio is 3 and the minimum -60. We use this information to normalize the Sharpe ratio. The ESG score is normalized considering its maximum and minimum values, which are 10 and 0. In order to analyse the effectiveness of the Bayesian optimization algorithm, we compare the results of the optimization with the best score achieved when running 100 different random variables for each of the three weights. Both the optimization and the random search are run for a budget of 25 iterations. We use the upper confidence bound acquisition function. For the Bayesian optimization to be effective, the average value, which is the expected value, of the fitness function has to be higher than for the random search.  

We propose two experiments with the same setting. In particular, we run Bayesian optimization and Random search 25 times with different random seeds. Every repetition of the experiments run 25 iterations of both methods. We hope that the expected value, approximated by the empirical mean, of Bayesian optimization is higher than the one of the Random Search method. Also, we hope that the standard deviation is lower in the case of Bayesian optimization, being a signal of a robust method.  

\begin{figure}[h]
    \centering
    \includegraphics[width = 0.99\textwidth]{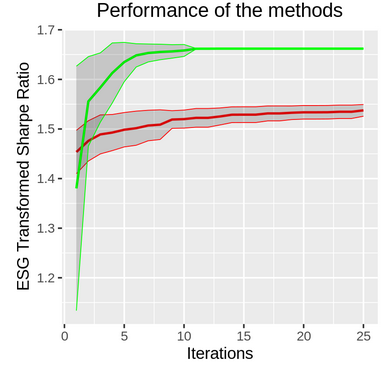}
\caption{Mean performance per iteration of Bayesian optimization (green) with respect to Random search (red) with respect to 25 different experiments. The standard deviation of the performance (ESG transformed Sharpe Ratio) is shown in grey for both methods.}
    \label{fig:im8}
\end{figure}

The first experiment involves the following ESG scores, for Endesa, 8.7, for Iberdrola, 8.97 and for Repsol, 7.32. In particular, we can see that the variability of these scores is not high according to its range.  Figure \ref{fig:im8} shows the mean performance of the best observed result in every iteration by Bayesian Optimization, in green, compared to the performance displayed by Random Search, in red.  

As it can be seen, Bayesian optimization outperforms Random Search, not only in performance but also in robustness, measured by the standard deviation on the mean performance. In particular, the best observed mean performance of Bayesian Optimization has been 1.662 and the performance of Random Search has been 1.538. Moreover, we can observe how Bayesian optimization repetitions all converge into an optimal portfolio.  

In particular, the optimum portfolio where all the repetitions of Bayesian optimization converge will be composed by 57,6\% shares of Endesa, 21,2\% shares of Iberdrola and 21,2\% shares of Repsol.  

We now give the details of a second experiment, where we set ESG scores with high variability. In particular, the new ESG scores are the following ones: [9,5,2]. We repeat again 25 experiments where every experiment includes 25 iterations of Bayesian Optimization and Random Search. Figure \ref{fig:im9} summarizes the obtained results: 

\begin{figure}[h]
    \centering
    \includegraphics[width = 0.99\textwidth]{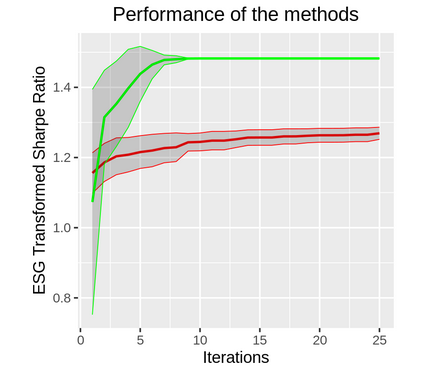}
\caption{Mean performance per iteration of Bayesian optimization (green) with respect to Random search (red) with respect to 25 different experiments in an experiment with ESG scores of high variability. The standard deviation of the performance (ESG transformed Sharpe Ratio) is shown in grey for both methods.}
    \label{fig:im9}
\end{figure}

We can see how, independently of the variability of the ESG scores, Bayesian Optimization outperforms Random Search, both in performance and robustness. Again, Bayesian Optimization repetitions converge into an optimal portfolio, whose ESG Transformed Sharpe Ratio is lower than in the case of the previous portfolio as a result of considering a lower sum of ESG scores, which penalizes the ESG Transformed Sharpe Ratio fitness function. 

Consequently, given the empirical evidence shown by the previous illustrative experiments, we can conclude that, for the ESG constrained portfolio optimization, Bayesian optimization could be used to design an optimal portfolio. 

\section{Conclusions and further work}
This paper has applied Bayesian optimization to an ESG constrained stock portfolio scenario. The optimization’s effectiveness has been proved in the illustrative experiment. The proposed method can be used to penalize the behaviour of a portfolio according to any criteria, such a Corporate Social Responsibility (CSR) criteria or a firm’s exposure to a certain country. For example, these penalizations could be used to take into account the consequences of being present in the Russian and Ukrainian market during the current war for multinational firms.   

Another interesting line of further research consists of decoupling the ESG constraints from the main objective and consider them as black-box constraints. Then, we will be able to avoid choosing at all costs different stock portfolio configurations using constrained multi-objective Bayesian optimization \cite{fernandez2020improved}. We would also like to compare the performance of Bayesian Optimization with respect to other black-box optimization methods as Tree Parzen Estimator and more complex Bayesian Optimization methodologies.

\bibliography{main}
\bibliographystyle{acm}

\end{document}